\documentstyle[epsfig]{mn}

\newif\ifAMStwofonts

\ifoldfss
  \ifCUPmtlplainloaded \else
    \NewTextAlphabet{textbfit} {cmbxti10} {}
    \NewTextAlphabet{textbfss} {cmssbx10} {}
    \NewMathAlphabet{mathbfit} {cmbxti10} {} 
    \NewMathAlphabet{mathbfss} {cmssbx10} {} 
  \fi
  \ifAMStwofonts
    \ifCUPmtlplainloaded \else
      \NewSymbolFont{upmath} {eurm10}
      \NewSymbolFont{AMSa} {msam10}
      \NewMathSymbol{\upi}     {0}{upmath}{19}
      \NewMathSymbol{\umu}     {0}{upmath}{16}
      \NewMathSymbol{\upartial}{0}{upmath}{40}
      \NewMathSymbol{\leqslant}{3}{AMSa}{36}
      \NewMathSymbol{\geqslant}{3}{AMSa}{3E}

    \fi
  \fi
\fi 

\ifnfssone
  \newmathalphabet{\mathit}
  \addtoversion{normal}{\mathit}{cmr}{m}{it}
  \addtoversion{bold}{\mathit}{cmr}{bx}{it}
  \newmathalphabet{\mathbfit} 
  \addtoversion{normal}{\mathbfit}{cmr}{bx}{it}
  \addtoversion{bold}{\mathbfit}{cmr}{bx}{it}
  \newmathalphabet{\mathbfss} 
  \addtoversion{normal}{\mathbfss}{cmss}{bx}{n}
  \addtoversion{bold}{\mathbfss}{cmss}{bx}{n}
  \ifAMStwofonts
    \ifCUPmtlplainloaded \else
      \UseAMStwoboldmath
      \makeatletter
      \new@mathgroup\upmath@group
      \define@mathgroup\mv@normal\upmath@group{eur}{m}{n}
      \define@mathgroup\mv@bold\upmath@group{eur}{b}{n}
      \edef\UPM{\hexnumber\upmath@group}
      \new@mathgroup\amsa@group
      \define@mathgroup\mv@normal\amsa@group{msa}{m}{n}
      \define@mathgroup\mv@bold\amsa@group{msa}{m}{n}
      \edef\AMSa{\hexnumber\amsa@group}
      \makeatother
      \mathchardef\upi="0\UPM19
      \mathchardef\umu="0\UPM16
      \mathchardef\upartial="0\UPM40
      \mathchardef\leqslant="3\AMSa36
      \mathchardef\geqslant="3\AMSa3E
    \fi
  \fi
\fi 

\ifnfsstwo
  \DeclareMathAlphabet{\mathbfit}{OT1}{cmr}{bx}{it}
  \SetMathAlphabet\mathbfit{bold}{OT1}{cmr}{bx}{it}
  \DeclareMathAlphabet{\mathbfss}{OT1}{cmss}{bx}{n}
  \SetMathAlphabet\mathbfss{bold}{OT1}{cmss}{bx}{n}
  \ifAMStwofonts
    \ifCUPmtlplainloaded \else
      \DeclareSymbolFont{UPM}{U}{eur}{m}{n}
      \SetSymbolFont{UPM}{bold}{U}{eur}{b}{n}
      \DeclareSymbolFont{AMSa}{U}{msa}{m}{n}
      \DeclareMathSymbol{\upi}{0}{UPM}{"19}
      \DeclareMathSymbol{\umu}{0}{UPM}{"16}
      \DeclareMathSymbol{\upartial}{0}{UPM}{"40}
      \DeclareMathSymbol{\leqslant}{3}{AMSa}{"36}
      \DeclareMathSymbol{\geqslant}{3}{AMSa}{"3E}
    \fi
  \fi
\fi 

\ifCUPmtlplainloaded \else
  \ifAMStwofonts \else 
    \def\upi{\pi}
    \def\umu{\mu}
    \def\upartial{\partial}
  \fi
\fi

\newcommand{\ea}{et~al.\ }
\title
[MEM reconstruction of gravitational lenses]
{A maximum-entropy method for reconstructing the projected mass 
distribution of gravitational lenses}
\author[S.L.~Bridle \ea]
{S.L.~Bridle, M.P.~Hobson, A.N.~Lasenby and Richard~Saunders}
\date{Accepted ???. Received ???; in original form \today}
\begin{document}
\maketitle
\begin{abstract} 
The maximum-entropy method is applied to the problem of reconstructing
the projected mass density of a galaxy cluster using its gravitational
lensing effects on background galaxies. We demonstrate the method by
reconstructing the mass distribution in a model cluster using
simulated shear and magnification data to which Gaussian noise is
added. The mass distribution is reconstructed directly and the
inversion is regularised using the entropic prior for this positive
additive distribution. For realistic noise levels, we find that
the method faithfully reproduces the main features of the cluster mass
distribution not only within the observed field but also slightly beyond it.
We estimate the uncertainties on
the reconstruction by calculating an analytic approximation
to the covariance matrix of the reconstruction values of each pixel.
This result is compared with error estimates derived from Monte-Carlo 
simulations for different noise realisations and found to be in good
agreement.
\end{abstract}

\section{Introduction}
\label{intro}

Galaxies and galaxy clusters can cause the images of more distant
galaxies to be distorted and magnified due to gravitational lensing
effects (see e.g. Schneider, Ehlers \& Falco 1992 or Blandford \&
Narayan 1992 for reviews). This naturally leads one to consider how we
may best use these effects to reconstruct the mass distribution of the
lensing cluster, particularly since such a reconstruction is sensitive
both to luminous and dark matter.

Typically the distances between the background objects and the lens,
and between the lens and the observer, are much larger than the size of
the lens itself so that the mass distribution of the lens can be
considered as a mass sheet lying the lens plane.  Thus a lens is
fully characterised by its surface mass density $\Sigma(\btheta)$,
which is a function of angular position $\btheta$ on the sky.

If at any point in the lens plane the surface mass density exceeds
some critical value $\Sigma_{\rm crit}$, which depends on the lensing
geometry (see Section~\ref{gravlens}), then the lens is said to be
supercritical and will exhibit non-linear `strong' lensing effects.
This condition is often satisfied by very dense clusters and numerous
examples of strong lensing have been observed (see e.g. Fort \&
Mellier 1994). Depending on the position of a background galaxy with
respect to the caustics of the lens, its image can be enormously
magnified and distorted so that it appears as a giant arc or is
multiply imaged. Such effects provide powerful constriants on the
projected mass in the lens that is contained within the arc.

In most cases, however, the gravitational lensing effects are more
subtle and typically the lensing cluster produces a large number of
weakly distorted images of background objects, which are called
arclets. 
The surface density of galaxies to an $R-$band magnitude of $R \sim
23$ mag is on the order of $\sim 10$ per square arcmin, rising to
$\sim 100$ per square arcmin for galaxies to $R \sim 25$ mag (e.g. Woods,
Fahlman \& Richer 1995). We therefore expect the typical separation of
arclets to be about 15 arcsec for observations out to $R \sim 23$ mag,
falling significantly to about 5 arcsec for $R \sim 25$ mag.
Since
we do not expect the gravitational potential of a cluster to change
appreciably over 5--10 arcsec scales, it is possible to average the
signals from several arclets at a time to produce a coherent pattern
of distortion. This averaging is usually performed by dividing the
image into square cells of size 30--60 arcsec, which will contain the
images of $\sim 10$--50 background galaxies.  The pattern of
distortion, or shear pattern, produced by the lensing cluster is
then measured by averaging the ellipticities and orientations of the
arclets in each cell; this is discussed further in
Section~\ref{simobs}.

The resulting shear pattern may then be used to estimate the surface
mass density $\Sigma(\btheta)$ of the lensing cluster. This procedure
was first investigated by Tyson, Valdes \& Wenk (1990) and
parameterised fits for cluster shapes were performed by Kochanek
(1990) and Miralda-Escud\'e (1991). The first parameter-free method
was presented by Kaiser \& Squires (1993) in which a Green's function
technique is used to reconstruct a two-dimensional map of
$\Sigma(\btheta)$ on the same grid of points as that used for
averaging ellipticities.

Although in its original form, the Kaiser \& Squires method applies
only in the weak lensing limit, it was extended to the non-linear
regime by Kaiser (1995). However, the method still has a number of
limitations, many of which are shared by other methods.  These
problems include
the `mass-sheet degeneracy' (Falco, Gorenstein \& Shapiro 1985;
Schneider \& Seitz 1995), which affects any algorithm that uses shear
information alone to determine the lens mass distribution; this can only
be broken by also measuring the magnification of background
galaxies.  Another problem faced by all reconstruction methods is 
a scaling uncertainty in the reconstructed surface mass
density if the redshifts of the lensed background objects are not
known, although some methods for circumventing this problem have been
proposed (e.g. Kneib et al. 1994; Smail, Ellis \& Fitchett 1994;
Bartelmann \& Narayan 1995).  Finally, perhaps the most important
drawback of the Kaiser \& Squires method is that it requires a
convolution of shears to be performed over the entire sky.  As a
result, if the field of observed shears is small or
irregularly-shaped, then the method can produce artefacts in the
reconstructed $\Sigma(\btheta)$ distribution near the boundaries of
the observed field. Nevertheless, several extensions of the basic
algorithm have been developed by Kaiser, Squires \& Broadhurst (1995);
Bartelmann (1995) and Seitz \& Schneider (1996).  These 
`finite-field methods' are based on a line-integral approach which
reduces the unwanted boundary effects.

More recently, Bartelmann et al. (1996) have presented a novel
maximum-likelihood approach to reconstructing the cluster mass
distribution that avoids boundary effects and uses both shear
and magnification data, thereby breaking the mass-sheet degeneracy.
The method is based on finding the Newtonian gravitational potential
that best reproduces the observed data in a straightforward least-squares
sense.  This best-fit gravitational potential is then used to obtain
the mass distribution. The advantage of this method is that, by
design, it can be applied directly to clusters where the observed
field is small or irregularly shaped. In addition, it is
straightforward to incorporate measurement inaccuracies and
correlations within the data.
Seitz, Schneider \& Bartelmann (1998) extend this approach to
include an entropy regularisation of the potential and use the
individual galaxy ellipticities and positions instead of averaging by
dividing the image into cells.
Squires \& Kaiser (1996) suggest several reconstruction techniques 
including two
maximum-likelihood methods. The most successful of these is
regularised in Fourier space using the prior that the
Fourier coefficients of the mass density distribution are drawn
from a Gaussian distribution of some constant width.

In this paper we consider maximum likelihood as a special case of the
maximum-entropy method (MEM) in the context of Bayes' theorem (see
Section~\ref{themem}). We use the MEM technique to reconstruct the
cluster mass distribution from a grid of 
simulated shear and magnification data
to which random Gaussian noise is added, 
although the basic method we describe could also be extended to use
individual galaxy ellipticities and positions.
We use the regularisation properties of MEM to reconstruct the mass distribution
directly rather than, for example, reconstructing the gravitational
potential as an intermediate step (Bartelmann et al. 1996, Seitz,
Schneider \& Bartelmann 1998), or assigning a prior to the Fourier
coefficients of the mass distribution (Squires \& Kaiser 1996).
This is clearly valuable since the
mass distribution is the quantity in which we are most interested and,
in addition, it allows a more straightforward evaluation of the
uncertainties in the reconstruction.  Moreover, since the mass
distribution is both positive and additive, it is natural to assign an
entropic prior. 
Without wanting to enter into the complex debate on this issue, there
are strong arguments which suggest that use of the entropy of the
image as the regularising function is the \emph{only} consistent way to
assign any positive additive distribution (see Gull 1989 for a review).
As a final point, we note that in a given observed
field, the distortions of the images of background galaxies are not
only influenced by the mass distribution inside the observed field but
also by mass lying outside the field. It is therefore necessary to
reconstruct the mass distribution some distance beyond the boundary of
the observed field, although the uncertainty in the reconstruction in
this region will clearly increase rapidly with distance since the
lensing effect is quite localised. As illustrated in
Section~\ref{application}, however, the MEM reconstruction of the mass
distribution remains faithful even slightly outside the observed
region. In regions far away the observed area, the entropic prior
ensures the reconstruction defaults to the assumed model value wherever
there is insufficient evidence in the data to the contrary.

\section{Gravitational lensing by a galaxy cluster}
\label{gravlens}

We begin by considering how the observed distortion and magnification of
the background galaxies depend on the mass distribution of the lensing
cluster. For a detailed account of gravitational lensing by clusters
see e.g. Schneider, Ehlers \& Falco (1992) or Blandford \& Narayan (1992);
here we summarise the points relevant to our current work.

\begin{figure}
\centerline{\epsfig{file=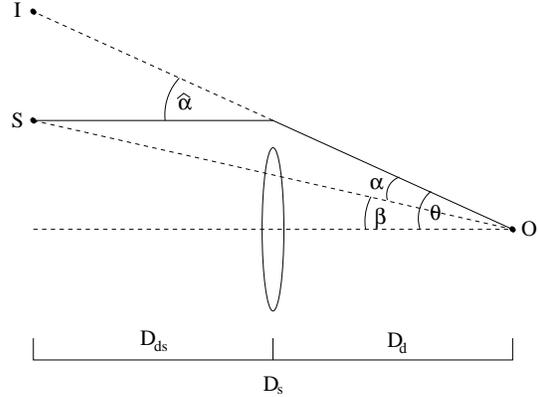,width=7cm}}
\caption{The gravitational lens geometry. The light ray propagates
from the source $S$ to the observer $O$ and is deflected through as
angle $\hat{\alpha}$ so that the image appears at $I$.
The angular separations of the source and image
from the optic axis are denoted by $\beta$ and $\theta$ respectively.
$D_{\rm{d}}$, $D_{\rm{s}}$ and $D_{\rm{ds}}$ are respectively the
angular-diameter distances from the observer to the lens, from the
observer to the source, and from the source to the lens.}
\label{fig1}
\end{figure}

The geometry assumed in this paper is illustrated in Fig.~\ref{fig1}.
Using Cartesian coordinates, the local properties of the lens mapping 
between source plane vectors $\bbeta=(\beta_1,\beta_2)$ and image plane vectors
$\btheta=(\theta_1,\theta_2)$ are given by the linear transformation
\begin{equation}
\rm{d}\bbeta = {\mathbfss A} \rm{d}\btheta,
\label{theta2beta}
\end{equation}
where the Jacobian matrix ${\mathbfss A}$ is the inverse of the magnification
tensor and is given by
\begin{equation}
\label{A}
\mathbfss{A}(\btheta) =
\left( \begin{array}{cc}
1-\kappa-\gamma_1 & -\gamma_2 \\
-\gamma_2 & 1-\kappa+\gamma_1
\end{array} \right).
\label{jacobian}
\end{equation}
The {\em convergence} $\kappa$ and the {\em shears} $\gamma_1$ 
and $\gamma_2$ are explained below. Note that in general the Jacobian
matrix is a function of position $\btheta$.

The convergence, $\kappa$, is obtained by rescaling the
projected mass per unit area in the lens, $\Sigma$, 
and is defined by
\begin{equation}
\kappa(\btheta)=\frac{\Sigma(\btheta)}{\Sigma_{\rm{crit}}},
\label{kappaSigma}
\end{equation}
where the critical surface density $\Sigma_{\rm crit}$ is given by
\begin{equation}
\Sigma_{\rm{crit}}=\frac{c^2}{4\pi G} 
\frac{D_{\rm{s}}} {D_{\rm{d}} D_{\rm{ds}}}.
\label{sigcrit}
\end{equation}
Here $D_{\rm{d}}$, $D_{\rm{s}}$ and $D_{\rm{ds}}$ are respectively the
angular-diameter distances from the observer to the lens, from the
observer to the source, and from the source to the lens; these
distances are shown in Fig.~\ref{fig1}.  Convergence acting alone
produces an isotropic magnification of background galaxies. 

The quantities $\gamma_1$ and $\gamma_2$ describe the anisotropic
distortion of background galaxy images. $\gamma_1$ describes the shear
in the $x$ and $y$ directions and $\gamma_2$ describes the shear in
the $x=y$ and $x=-y$ directions.
For algebraic
simplicity they are often combined into the complex shear
\begin{equation}
\gamma(\btheta)=\gamma_1(\btheta)+ \rm{i}\gamma_2(\btheta).
\label{compshear}
\end{equation}
The modulus $|\gamma|$ describes the
magnitude of the shear and its argument, $\phi$, describes the
orientation.
In the presence of both convergence and shear, a circular source of
unit radius becomes an elliptical image with major and minor axes of
lengths 
\begin{equation}
a=(1-\kappa-|\gamma|)^{-1},\hspace{1cm}b=(1-\kappa+|\gamma|)^{-1}.
\label{majmin}
\end{equation}
and the total magnification is given by
\begin{equation}
\mu=\frac{1}{{\rm det}{\mathbfss A}} = \frac{1}{(1-\kappa)^2-|\gamma|^2}.
\label{magnif}
\end{equation}

The shear $\gamma$ and the convergence $\kappa$ are related
by the scaled two-dimensional gravitational
potential $\psi$ (see Schneider, Ehlers \& Falco 1992). Both $\gamma$ and
$\kappa$ can be written as linear combinations of this potential and
are given by
\begin{eqnarray}
\gamma_1(\btheta) &=& {\textstyle\frac{1}{2}} \left( \psi_{,11}(\btheta) 
- \psi_{,22}(\btheta) \right),
\label{gamma1pot}\\
\gamma_2(\btheta) &=& \psi_{,12}(\btheta),
\label{gamma2pot} \\
\kappa(\btheta) &=& {\textstyle\frac{1}{2}} \left( \psi_{,11}(\btheta) 
+ \psi_{,22}(\btheta) \right),
\label{kappapot}
\end{eqnarray} 
where the commas and subscripts on $\psi$ denote partial differentiation with
respect to the image plane vectors $\theta_i$, for example,
$\psi_{,12}\equiv\partial^2\psi/\partial\theta_1\partial\theta_2$. 
Indeed the relationships (\ref{gamma1pot})--(\ref{kappapot}) form the
basis of the maximum-likelihood
reconstruction method presented by Bartelmann et al. (1996) in which
the observed distortion and magnification of background galaxies are used to
reconstruct the gravitational potential of the lensing cluster.
The corresponding mass distribution is then found using
(\ref{kappapot}), (\ref{kappaSigma}) and (\ref{sigcrit}).

\section{Simulating observations of gravitational lensing}
\label{simobs}

In the approach presented below we use shear
and magnification data to reconstruct the
convergence $\kappa(\btheta)$, which can then be converted
into the surface mass density $\Sigma(\btheta)$ using (\ref{kappaSigma}).
In order to achieve this using our method, it is first necessary 
to consider the
`forward' problem of predicting the shear and magnification pattern
produced by a given convergence distribution.

Fourier transforming equations (\ref{gamma1pot})--(\ref{kappapot}), we find
\begin{eqnarray}
\tilde{\gamma}_1(\bmath{k}) &=& 
-{\textstyle\frac{1}{2}} \left(k_1^2 - k_2^2\right)
\tilde{\psi}(\bmath{k}),   \\
\tilde{\gamma}_2(\bmath{k}) &=& 
-k_1 k_2 \tilde{\psi}(\bmath{k}) \\
\tilde{\kappa}(\bmath{k}) &=& 
-{\textstyle\frac{1}{2}} \left(k_1^2 + k_2^2\right)
\tilde{\psi}(\bmath{k})
\end{eqnarray}
where $\bmath{k}=(k_1,k_2)$ is the position vector in Fourier space.
Eliminating $\tilde{\psi}$ from these equations we obtain
\begin{eqnarray}
\tilde{\gamma}_1(\bmath{k}) 
&=& \frac{\left(k_1^2 - k_2^2\right)}{|\bmath{k}|^2}
\tilde{\kappa}(\bmath{k}),   \\
\tilde{\gamma}_2(\bmath{k}) &=& \frac{2 k_1 k_2}{|\bmath{k}|^2} 
\tilde{\kappa}(\bmath{k}).
\end{eqnarray}
Inverse Fourier transforming and
combining the shear components $\gamma_1$ and $\gamma_2$ into the
complex shear $\gamma$, we find
\begin{equation}
\gamma(\btheta)=\frac{1}{\pi} \int  
{\mathcal D}(\btheta- \btheta')\kappa(\btheta')\,\rm{d}^2 \btheta',
\label{gammadef}
\end{equation}
where the kernel is given by
\begin{equation}
{\mathcal D}(\btheta)=\frac{\theta_2^2-\theta_1^2- 2\rm{i}\theta_1 \theta_2}
{ |\btheta|^4}.
\label{kernel}
\end{equation}
Thus, equations (\ref{gammadef}) and (\ref{kernel}) express the shear
pattern as a convolution in terms of the convergence $\kappa$. For a
given $\kappa$ distribution it is then also straightforward to obtain
the resulting magnification pattern $\mu(\btheta)$ using
(\ref{magnif}).

As mentioned in Section~\ref{intro}, lensing data typically consist of
an observed field made up of a grid of cells 
in each of which the average distortion and magnification of the
background sources have been calculated.  In each cell the average
distortion can be quantified by measuring the mean ellipticity of the
lensed images in that cell, whereas the magnification can be found
either by comparing galaxy counts in the cluster field and in an
unlensed field (Broadhurst, Taylor \& Peacock 1995; Seitz \& Schneider
1997), or by comparing
the sizes of similiar galaxies in two such fields (Bartelmann \&
Narayan 1995).

To produce realistic simulated observations of the lensing induced by
a given model cluster, it is usual to assume some population of
background galaxies (see Bartelmann et al. 1996) and compute
numerically the resulting arclets after lensing by the model cluster.
These arclets are then analysed to produce a grid of averaged
ellipticites and magnifications, as discussed above. This technique
also allows an estimate to be made of the uncertainty in the
determination of the averaged quantities in each cell.

\begin{figure}
\centerline{\epsfig{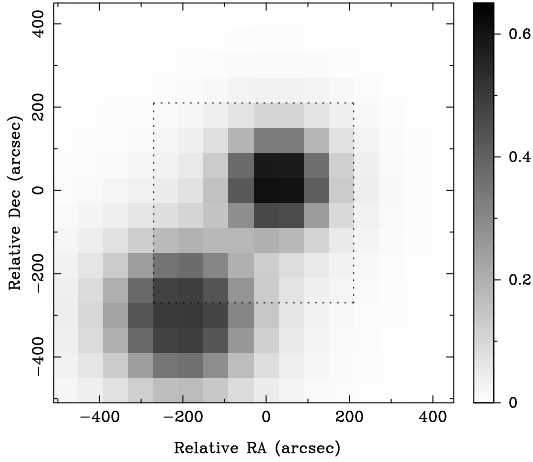}}
\caption{The true convergence distribution $\kappa(\btheta)=
\Sigma(\btheta)/\Sigma_{\rm crit}$ of the model cluster.
The dotted line indicates the edge of the observed field.}
\label{fig2}
\end{figure}

In order to assess the capabilities of the MEM reconstruction
technique, we adopt an equivalent approach in our simulations,
although in a slightly more straightforward manner.  To simulate an
observation, we begin by assuming a convergence distribution
$\kappa(\btheta)$ for our model cluster.  The model cluster used is
shown in Fig.~\ref{fig2} and consists simply of two isothermal
spheres projected onto the lens plane.
The cluster is simulated on a $16\times 16$ grid with a cell size of 1
arcmin and is everywhere subcritical. We then assume that the observed
field consists only of the $8 \times 8$ grid cells at the centre of
the $16 \times 16$ grid; this is delineated by the dashed box in
Fig.~\ref{fig2}.

In each cell of the observed field the average ellipticities and
magnifications, in the absence of noise, are then calculated as
follows. The complex ellipticity of an image is defined as
\[
\epsilon = \epsilon_1+\mbox{i}\epsilon_2= \frac{a-b}{a+b}\exp(i\phi).
\]
where $a$ and $b$ are respectively the lengths of the major and minor
axes of the elliptical image and $\phi$ is its position angle with
respect to the $x$-axis. If we assume that the intrinsic ellipticities
of the background sources within a cell average to zero, then using
(\ref{majmin}) the predicted average ellipticities
$\epsilon^{\rm (p)}_i$ $(i=1,2)$ in the cell due to lensing by the
cluster are given simply by the mean values of $\gamma_i/(1-\kappa)$
over the cell, where $\gamma_i$ is calculated using (\ref{gammadef}).
The predicted average magnification $\mu^{\rm (p)}$ in each grid
cell can be calculated in a similar way, but in practice it is more
convenient to work in terms of its inverse $r^{\rm (p)}$, which
from (\ref{magnif}) is given simply by the average of
$(1-\kappa)^2-|\gamma|^2$ over the cell.

Finally, random Gaussian noise is added to the predicted average
ellipticities and inverse magnifications in each cell. For the
ellipticities, the standard deviation of the noise on each grid cell
is assumed to be $\sigma_\epsilon = 0.05$ and for the inverse
magnifications we assume $\sigma_r = 0.1$.  The value $\sigma_\epsilon
= 0.05$ is consistent with that expected for 20 galaxies per grid cell
(Schneider \& Seitz 1995).  If found from galaxy counts,
the estimate of the magnification will be subject to Poisson noise,
and for 30 galaxies per cell the error will be of the order of 10 per
cent, and the noise on the inverse magnification will also be of the
same size. Thus our assumption of an error $\sigma_r=0.1$ on the
inverse magnification is not unreasonable.
\begin{figure*}
\centerline{\epsfig{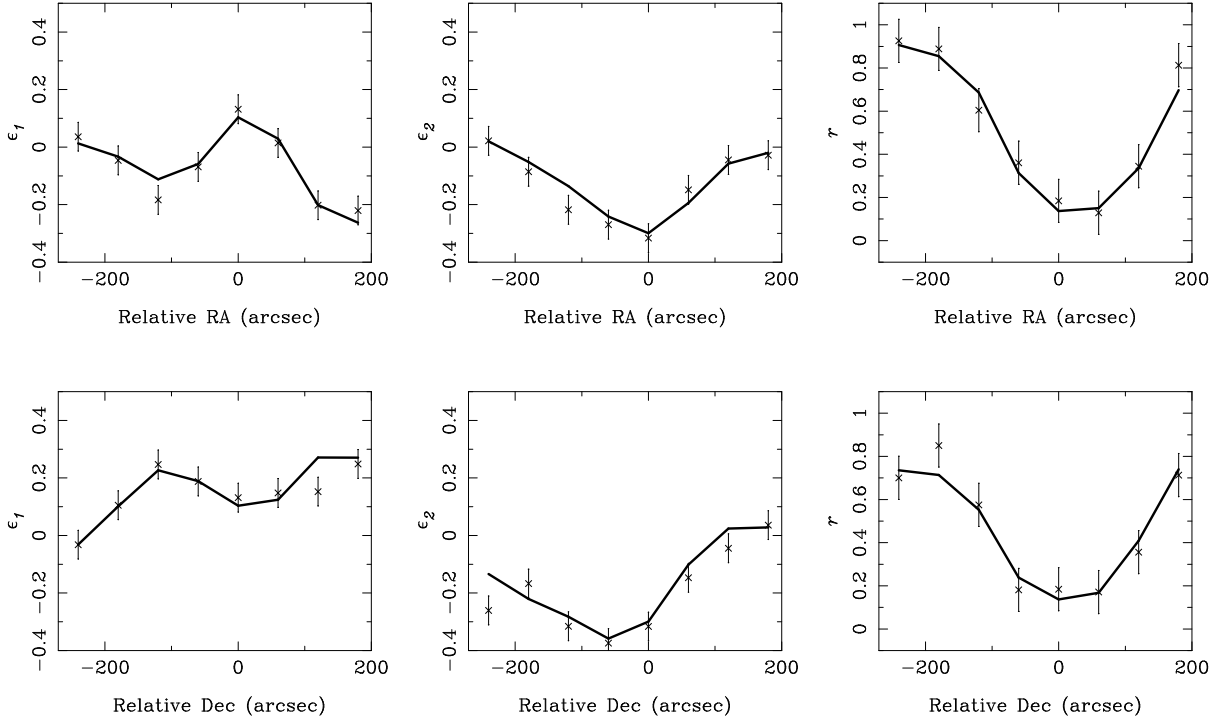}}
\caption{Slices along the $x$ and $y$-axes for each of the observed
quantities, $\epsilon_1$, $\epsilon_2$ and $r$. The continuous line is
the quantity calculated directly from the mass distribution, with no
noise added. The points are observed values of each quantity for a
particular noise realisation.  The error bars indicate the noise
levels used.}
\label{fig3}
\end{figure*}

Thus the basic data in our simulated lensing observation
consist of, in each grid cell, the `observed' average ellipticities 
$\epsilon^{\rm (o)}_1$ and $\epsilon^{\rm (o)}_2$ 
and inverse magnification $r^{\rm (o)}$, each of which contains
a noise contribution.
As an illustration of the simulated lensing data, in
Fig.~\ref{fig3} we plot the average ellipicity and inverse
magnification produced by our model cluster along the $x$ and $y$-axes
in Fig.~\ref{fig2}. The solid line represents the value of each
quantity in the absence of noise, the points denote the observed
values for our given noise realisation and the error bars indicate
the rms noise level assumed.

\section{The maximum-entropy method}
\label{themem}

As explained in the previous section, the simulated `observed' data 
consist of the three quantities $\epsilon^{\rm (o)}_1$, 
$\epsilon^{\rm (o)}_2$ and
$r^{\rm (o)}$ in each grid cell of the observed field.
Let us denote these observed data by the vector 
\[
{\mathbfss d}_{\rm o}
=(\epsilon^{\rm (o)}_1(1),\epsilon_2^{\rm (o)}(1),r^{\rm (o)}(1),\ldots,
\epsilon_1^{\rm (o)}(N),\epsilon_2^{\rm (o)}(N),r^{\rm (o)}(N)),
\]
where $\epsilon_1^{\rm (o)}(j)$ and $\epsilon_2^{\rm (o)}(j)$ denote the
observed components of the average ellipticity in the $j$th grid cell
and $r^{\rm (o)}(j)$ denotes the corresponding observed inverse
magnification.  The total number of observed grid cells is $N=8\times
8=64$ in our simulation and thus the vector ${\mathbfss d}_{\rm o}$
has $3N = 192$ components.

From these data, we wish to
reconstruct the cluster mass distribution (defined in terms of 
the convergence $\kappa$)
on some grid of points. For simplicity we use the same grid for the 
reconstruction as that on which the cluster mass distribution was 
originally defined (see Fig.~\ref{fig2}), although this is not
required by the algorithm. Hence we aim to reconstruct the convergence
digitised on to $16 \times 16$ cells and we denote this distribution
by the vector 
\[
\bkappa=(\kappa(1),\kappa(2),\ldots,\kappa(L)),
\]
where $\kappa(j)$ is the convergence in the $j$th cell and $L$ is the 
total number of pixels in the reconstruction grid (i.e. $16 \times 16$).
Since the observed field consists only of 
the central $8 \times 8$ grid cells, we are therefore attempting
to reconstruct the cluster mass distribution some way outside the
region in which data are available.

We choose our estimator $\hat{\bkappa}$ 
of the convergence distribution to be that which maximises
the posterior probability $\Pr(\bkappa|{\mathbfss d}_{\rm o})$. 
Using Bayes' theorem, this is given by
\begin{equation}
\Pr(\bkappa|{\mathbfss d}_{\rm o}) 
= \frac{\Pr({\mathbfss d}_{\rm o}|\bkappa)\Pr(\bkappa)}
{\Pr({\mathbfss d}_{\rm o})},
\end{equation}
where $\Pr({\mathbfss d}_{\rm o}|\bkappa)$ is the likelihood of obtaining
the data given a particular convergence distribution and
the evidence $\Pr({\mathbfss d}_{\rm o})$ is simply a constant
that ensures
the posterior probability is correctly normalised. The prior
probability $\Pr(\bkappa)$ codifies our expectations about
the convergence distribution before acquiring the data 
${\mathbfss d}_{\rm o}$. Since the evidence is merely normalisation
factor, we actually need only to maximise the product
$\Pr({\mathbfss d}_{\rm o}|\bkappa)\Pr(\bkappa)$.

Let us first consider the form of the likelihood 
$\Pr({\mathbfss d}_{\rm o}|\bkappa)$. 
As discussed above, given a convergence distribution
$\bkappa$ we can use equations (\ref{gammadef}) and 
(\ref{magnif})
to predict the values of the noiseless average ellipticities and inverse
magnification in each observed grid cell. We denote these 
predicted values in the $j$th grid cell
as $\epsilon_1^{\rm (p)}(j)$, $\epsilon_2^{\rm (p)}(j)$ 
and $r^{\rm (p)}(j)$ and assemble the values for all the cells into
the predicted data vector
\[
{\mathbfss d}_{\rm p}
=(\epsilon^{\rm (p)}_1(1),\epsilon_2^{\rm (p)}(1),r^{\rm (p)}(1),\ldots,
\epsilon_1^{\rm (p)}(N),\epsilon_2^{\rm (p)}(N),r^{\rm (p)}(N)).
\]
Therefore, we may consider the observed data vector as
\[
{\mathbfss d}_{\rm o} = {\mathbfss d}_{\rm p} + {\mathbfss n},
\]
where the vector ${\mathbfss n}$ contains the noise contribution
in each observed grid cell for our particular realisation. 

If the noise on the
observed data values is Gaussian-distributed 
then the likelihood function is the $3N$-dimensional multivariate
Gaussian 
\[
\Pr({\mathbfss d}_{\rm o}|\bkappa)
\propto 
\exp\left[-{\textstyle\frac{1}{2}} 
({\mathbfss d}_{\rm o}-{\mathbfss d}_{\rm p})^{\rm T}
{\mathbfss N}^{-1}({\mathbfss d}_{\rm o}-{\mathbfss d}_{\rm p})\right],
\]
where ${\mathbfss N}=\langle {\mathbfss nn}^{\rm T}\rangle$ is the
ensemble average noise covariance matrix. If there are any
correlations between the different measured quantities, either within
a cell or between cells, then they can be
straightforwardly incorporated into the noise covariance matrix.
For example, Bartelmann et al. (1996) point out that
for faint lensed images it is possible that the measured values of the
ellipticities and magnification can be correlated. In most cases,
however, it is assumed that the data values are all independent, 
so the noise covariance matrix becomes diagonal and is given by
\[
{\mathbfss N}
=\mbox{diag}(\sigma^2_{d(1)},\ldots,\sigma^2_{d(3N)}),
\]
where $\sigma^2_{d(j)}$ is the variance of the noise in the
$j$th data value. Then the likelihood becomes
\begin{equation}
\Pr({\mathbfss d}_{\rm o}|\bkappa) 
\propto \exp\left(-{\textstyle\frac{1}{2}}\chi^2\right),
\label{likelihood}
\end{equation}
where $\chi^2$ is the standard misfit statistic given by
\begin{equation}
\chi^2 = \sum_{j=1}^{3N} \frac{\left[d_{\rm o}(j)-d_{\rm p}(j)\right]^2}
{\sigma^2_{d(j)}}.
\label{chisqdef}
\end{equation}

Let us now turn our attention to the prior probability
$\Pr(\bkappa)$. The simplest possible prior is the uniform
prior, which assumes that in each grid cell, before acquiring any data,
all values of the convergence are equally likely.
In this case the posterior probability is
directly proportional to the likelihood and so by maximising
the posterior with respect to $\bkappa$, we actually obtain the 
maximum-likelihood estimator for the convergence distribution.
However, since the convergence is a positive additive 
distribution, it may be shown (Skilling 1989),
using very general notions of subset invariance,
coordinate invariance and system independence, 
that the prior probability assigned to the components
of the vector $\bkappa$ should take the form
\begin{equation}
\Pr(\bkappa) \propto \exp[\alpha S(\bkappa,{\mathbfss m})],
\label{entprior}
\end{equation}
where the dimensional constant $\alpha$ depends on the scaling of
the problem and may be considered as a regularising parameter, and
${\mathbfss m}$ is a model vector to which $\bkappa$ defaults in the
absence of any evidence in the data to the contrary. The function
$S(\bkappa,{\mathbfss m})$ is the {\em cross-entropy} of 
$\bkappa$ and ${\mathbfss m}$ and is given by
\begin{equation}
S(\bkappa,{\mathbfss m}) = \sum_{j=1}^L
\kappa(j)-m(j)-\kappa(j)\ln\left(\frac{\kappa(j)}{m(j)}\right),
\label{entdef}
\end{equation}
which has a global maximum at $\bkappa={\mathbfss m}$.
If we have some prior knowledge of the structure of the lensing
cluster, we may include this information in the model $\mathbfss{m}$.
Otherwise, we set the model to have the same value in each pixel.
This value is generally set to be somewhat smaller than the
level expected for the reconstructed convergence distribution, but
in regions where the convergence is well-constrained by the data, the
level of the model makes no difference to the final reconstruction
(see Section~\ref{application}).

Combining the expression (\ref{likelihood}) and (\ref{entprior})
for the likelihood
and prior respectively, we see that the posterior probability
distribution is given by
\begin{equation}
\Pr(\bkappa|{\mathbfss d}_{\rm o}) \propto
\exp(-{\textstyle\frac{1}{2}}\chi^2+\alpha S).
\label{postdef}
\end{equation}
Therefore, maximising this distribution with respect to
$\bkappa$ is equivalent to minimising the function
\begin{equation}
F={\textstyle\frac{1}{2}}\chi^2-\alpha S.
\label{objfun}
\end{equation}

From the expressions (\ref{chisqdef}) and (\ref{entdef}), it is possible to
calculate analytically the first and second partial derivatives of $F$
with respect to $\kappa(j)$ ($j=1,2,\ldots,L$).
In order to minimise the function $F$ we use a conjugate
gradient algorithm in which the analytical derivative calculations are
performed using Fast Fourier Transforms.  Specifically, we use the
Polak-Ribiere variant of Fletcher-Reeves conjugate gradient method
(Press \ea 1992), with modifications to the implementation as
suggested by Mackay (1996). In fact, since the convergence distribution is
always positive it is convenient to minimise with respect to 
$\log\kappa(j)$ for which the corresponding derivatives of $F$
are also easily found.

Finally, we must consider the value of the regularising
parameter $\alpha$. In early implementations of MEM, $\alpha$ was chosen
so that for the final reconstruction the misfit statistic $\chi^2$ 
equalled its expectation
value, i.e. the number of data values $3N$. This choice is usually refered
to as historic MEM. It is, however, possible to determine the
appropriate value for $\alpha$ in a fully Bayesian manner 
(Skilling 1989; Gull \& Skilling 1990) by simply treating it as
another parameter in our hypothesis space. It may be shown that
$\alpha$ must satisfy
\begin{equation}
-2\alpha S(\hat{\bkappa},{\mathbfss m})
=L-\alpha\mbox{Tr}({\mathbfss M}^{-1}),
\label{bayesalpha}
\end{equation}
where $\hat{\bkappa}$ is the convergence distribution that minimises
the function $F$ for this value of $\alpha$ and $L$ is the total
number of pixels in the reconstruction. The $L \times L$ matrix
${\mathbfss M}$ is given by
\[
{\mathbfss M}={\mathbfss G}^{-1/2}{\mathbfss H}{\mathbfss G}^{-1/2},
\]
where ${\mathbfss H}=\nabla_{\bkappa}\nabla_{\bkappa}F$ is the Hessian
matrix of the function $F$ evaluated at the point $\hat{\bkappa}$ 
and ${\mathbfss G}=-\nabla_{\bkappa}\nabla_{\bkappa}S$ is minus the 
Hessian matrix of the entropy function at $\hat{\bkappa}$.
As mentioned above, these matrices can be calculated analytically.

It may also be shown, however, that the Bayesian value for
$\alpha$ may be reasonably well approximated by simply choosing
its value such that the value of $F$ at its minimum
is equal to the half number of data points, i.e.
$F(\hat{\bkappa})=3N/2$ (see Mackay 1996). We have determined
$\alpha$ by both methods and found them to agree to within a
few percent. More importantly, the resulting reconstructions of the
convergence distributions for each case are indistinguishable.

\section{Estimating the reconstruction errors}
\label{esterrors}

It is essential to be able to estimate the errors in the
reconstruction. From (\ref{postdef}) and (\ref{objfun}), we see that
the posterior probability distribution may be written
\[
\Pr(\bkappa|{\mathbfss d}_{\rm o}) \propto \exp[-F(\bkappa)].
\]
In general, the posterior distribution may 
have a complicated shape. Nevertheless, we may approximate its shape
near the peak $\hat{\bkappa}$ by a Gaussian of the form
\begin{equation}
\Pr(\bkappa|{\mathbfss d}_{\rm o}) 
\propto \exp\left[-{\textstyle\frac{1}{2}}(\bkappa-\hat{\bkappa})^{\rm T}
{\mathbfss H}(\bkappa-\hat{\bkappa})\right],
\label{gaussapp}
\end{equation}
where ${\mathbfss H}$ is the Hessian matrix mentioned above.
Comparing (\ref{gaussapp}) with the standard 
form for a multivariate Gaussian distribution, we see that the
covariance matrix of the errors on the $\bkappa$ reconstruction are
given approximately by the inverse of the Hessian matrix, i.e.
%
\begin{equation}
\langle (\bkappa-\hat{\bkappa})(\bkappa-\hat{\bkappa})^{\rm T}\rangle
\approx {\mathbfss H}^{-1},
\end{equation}
%
where the angle brackets denote the ensemble average.
The Hessian matrix at the peak of the posterior distribution
can be calculated analytically and has dimension $L\times L$,
where $L$ is the total number of pixels in the reconstruction.
In our simulation $L=256$, and so the numerical inversion of the
matrix may be performed using standard techniques in
just a few seconds of CPU time. This therefore provides 
a quick approximation
to the covariance matrix of the reconstruction errors.

In order for the above method to be reliable, we require the Gaussian
approximation at the peak of the posterior distribution to be
reasonably accurate. An alternative method for estimating the errors
on the reconstruction, which does not rely on any approximations, is
to use Monte-Carlo simulations. By repeating the analysis of
the same lensing data many times, but using different noise
realisations, we can measure the covariances of the 
reconstruction errors numerically. However, this technique only
provides an estimate of the ensemble average errors, rather than the
errors associated with the reconstruction obtained
from particular noise realisation. In addition, this approach is
more computationally intensive.

\section{Application to simulated observations}
\label{application}

\begin{figure}
\centerline{\epsfig{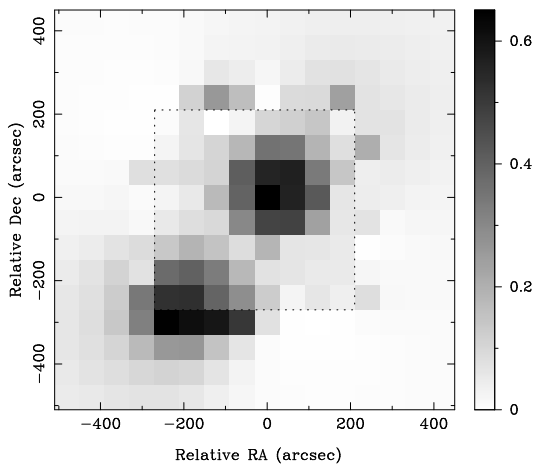}}
\caption{The maximum-entropy reconstruction of the convergence distribution
of the model cluster shown in Fig.~\ref{fig2}. The dotted line
indicates the edge of the observed field.}
\label{fig4}
\end{figure}

We have applied the MEM approach to the simulated ellipticity and
magnification data discussed in Section~\ref{simobs}, which were
produced assuming the cluster convergence distribution shown in
Fig.~\ref{fig2}.  Assuming the model $\mathbfss{m}$ in
(\ref{entdef}) to have a uniform value of 0.02 across the whole field,
the resulting MEM reconstruction of the convergence distribution is
shown in Fig.~\ref{fig4}. This reconstruction converges after about
90 iterations of the conjugate gradient algorithm to the point where the
fractional change in $F$ per iteration is less that 0.1 per cent.
The total CPU time required is about 2 minutes on a Sparc Ultra
workstation. The Bayesian value of the regularising parameter $\alpha$
that satisfies (\ref{bayesalpha}) was found to be $\alpha=2.6$,
On performing a further 100 iterations to achieve an even
more tightly defined convergence, the fractional change in $F$ per
iteration decreases to only 0.001 per cent and, more importantly, this
makes no discernable difference to the reconstruction. Similar
performances were found for numerous other noise realisations and we are
therefore confident that the MEM solution is extremely stable.

Comparing the reconstruction with the true distribution shown in
Fig.~\ref{fig2}, we see that within the observed field (delineated
by the dashed box), the MEM reconstruction has faithfully reproduced the
main features. In the first layer of pixels outside the box some
spurious features occur but, as discussed below, these
are easily shown not to be significant. The algorithm
succeeded in reconstructing the mass condensation
to the bottom left of the observed field. This feature
in the reconstruction is in fact found to be significant, despite
the fact that much of it lies outside the observing box.  We also note that the
entropic regularisation of the reconstruction has resulted in the
outer edges defaulting to the model value in the absence of any data
to the contrary. 

\begin{figure}
\centerline{\epsfig{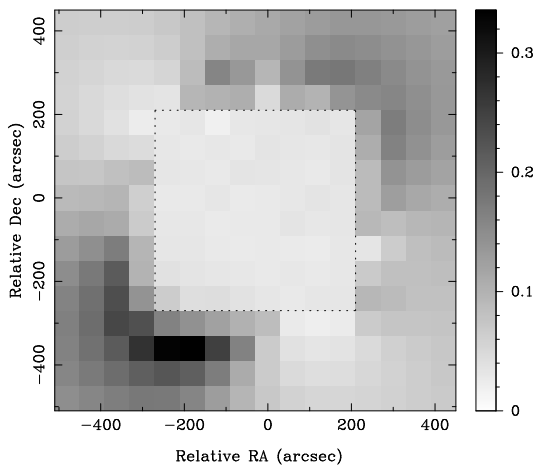}}
\caption{The rms errors in the MEM reconstruction of the cluster
convergence, as estimated by making a Gaussian approximation to the
posterior probability distribution at its peak. The dotted line
indicates the edge of the observed field.}
\label{fig5}
\end{figure}

One may investigate the accuracy of the reconstruction and the
significance of various features using either of the techniques
discussed in Section~\ref{esterrors}. 
First, we use the Gaussian approximation to the
posterior probability distribution to estimate the covariance matrix
of the reconstruction errors. The square-root of the diagonal elements
of this matrix give the rms errors in the reconstruction in each grid
cell, and these are plotted in Fig.~\ref{fig5}. Comparing
this with the MEM reconstruction shown in Fig.~\ref{fig4} and the true
convergence distribution in Fig.~\ref{fig2}, we see that the rms errors are
uniformly low within the observing box, with a typical value of about
0.05. Outside the observing box the rms errors increase, as expected.
In particular, we see that rms errors are indeed highest in the grid cells
where the reconstruction differs most noticeable from the true distribution.

\begin{figure*}
\centerline{\epsfig{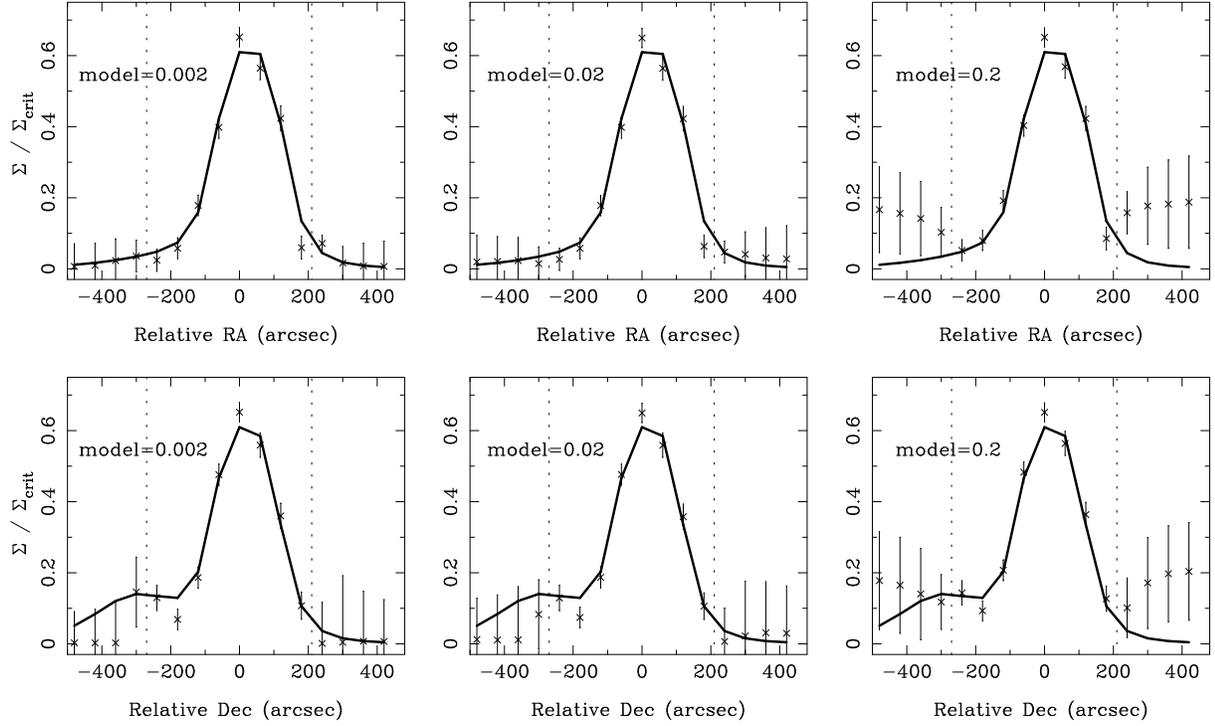}}
\caption{Slices through the $x$ and $y$ axes of the convergence distribution.
The continuous line is the true distribution and the points are 
the reconstructed values. The error bars show the standard deviation 
of the reconstruction errors in each cell, as estimated by making a
Gaussian approximation to the peak of the posterior probability
distribution.}
\label{fig6}
\end{figure*}

As a further illustration of the accuracy of the reconstruction, 
in Fig.~\ref{fig6} we plot slices along the $x$ and $y$-axes
for the true convergence distribution (solid line) and the
MEM reconstruction (points); we also plot the rms errors 
on the reconstruction predicted
by the Gaussian approximation to the posterior distribution. 
In addition to reconstruction discussed above, for which the
level of the flat model ${\mathbfss m}$
was set to 0.02, Fig.~\ref{fig6} also contains similar plots for
reconstructions calculated from the same lensing data, but for which
the models levels were set to 0.002 and 0.2 respectively. These
plots show that within the observing box the level of the model
makes virtually no difference to the reconstruction. Outside the
observing box, we see that, as expected, the reconstruction tends to the
model value in regions where there is no evidence in the data to the 
contrary. More importantly, we see that in all the plots shown 
in Fig.~\ref{fig6} the difference
between the true and reconstructed distributions is consistent with
the calculated error bars.

\begin{figure*}
\centerline{\epsfig{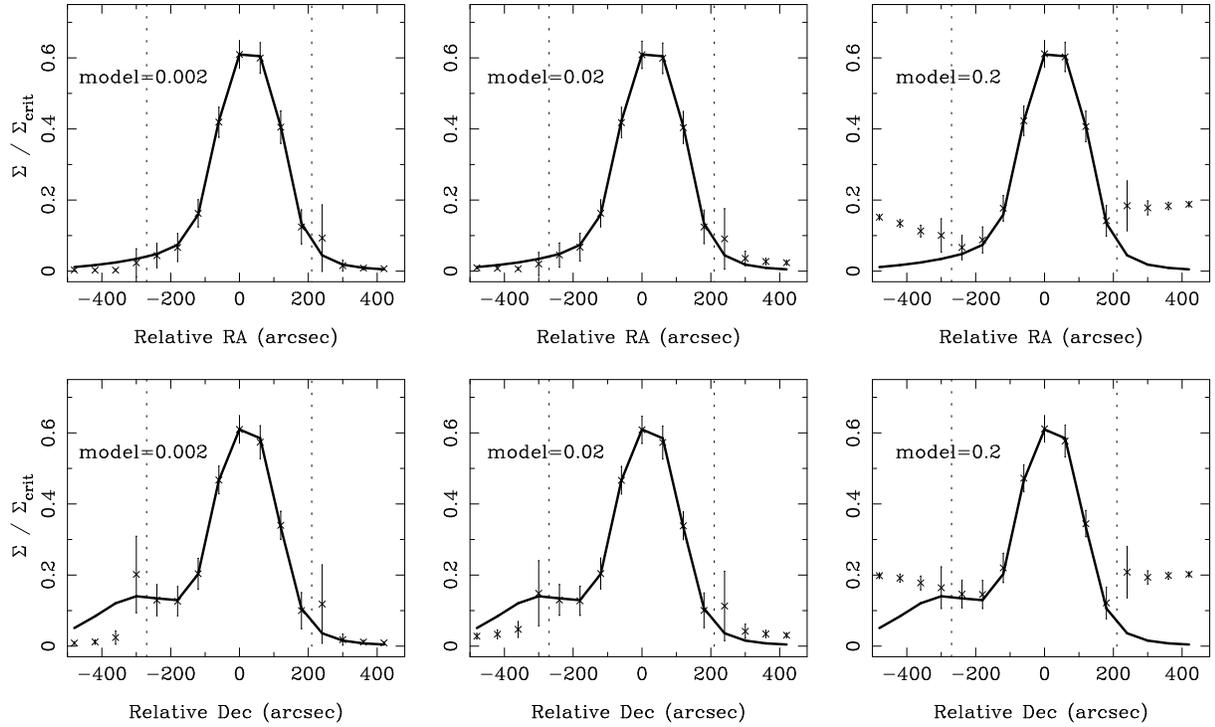}}
\caption{Slices through the $x$ and $y$ axes of the convergence distribution.
The continuous line is the true distribution and the points are the mean 
values of the convergence in each cell averaged over 100 noise realisations.
The error bars show
the standard deviations of the convergence in each cell calculated
directly from the 100 reconstructions using different noise realisations.}
\label{fig7}
\end{figure*}

As mentioned in Section~\ref{esterrors}, we may also investigate the accuracy
of the MEM reconstruction by performing Monte-Carlo simluations in
which the same lensing data are analysed but for different
noise realisations. In Fig.~\ref{fig7} we plot the results
obtained for
100 Monte-Carlo realisations, in the same format as Fig.~\ref{fig6}.
The points denote the mean reconstructed convergence obtained in each
grid cell averaged over the 100 realisations.
The errors bars denote the standard deviation of the convergence
in each grid, as measured directly from the realisations.
In regions where the convergence distribution 
is well-determined by the observations, we see that the mean reconstructed
value is very close to the true distribution, indicating that there no
bias in the MEM reconstruction. Moreover, in the observed field,
the sizes of the error bars agree closely with those in
Fig.~\ref{fig6}, which shows that the Gaussian approximation to the
posterior probability distribution is quite accurate.
In regions where
the data do not constrain the convergence, we see that the
reconstruction always defaults to the assumed model level and hence
the ensemble-average error bars are very small. 


\begin{figure*}
\centerline{\epsfig{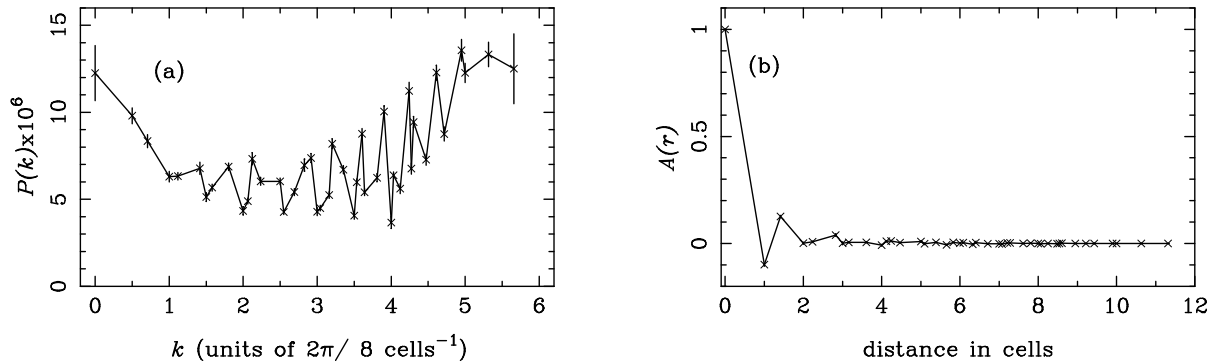}}
\caption{The azimuthally averaged power spectrum (a) and
autocorrelation function (b) of the
residuals (reconstructed convergence minus true convergence)
in the observed field averaged over 100 noise realisations. The error
bars show the 1--sigma uncertainties in each point as derived from the
100 realisations. The plotted power spectrum is the azimuthal
average of the 2D power spectrum of the residuals map. The
normalisation was taken such that the sum of
the points in the 2D power spectrum equalled the mean squared value of
the residuals map. The normalisation of the autocorrelation function
was fixed so that it equalled unity at zero separation.
}
\label{fig8}
\end{figure*}

In addition we have carried out a `noise power analysis' like that of
Seitz \& Schneider (1996) and Squires \& Kaiser (1996). In Fig.
\ref{fig8} (a) we plot the 
azimuthal average of the 2D
power spectrum of the residuals between
the reconstructed field and the original field (taken within the
observed field only), averaged over 100 noise realisations. 
The $k=0$ power point is due to the square of the mean of the
residual map, averaged over realisations. However, apart from this
point,
the power is low for small $k$ and increases for large $k$. This reflects
the fact that the uncertainties in the mass densities of neighbouring
pixels are strongly correlated, but the errors in widely spaced pixels
are not. This is more easily seen in the auto correlation function of
the residual map plotted in Fig. \ref{fig8} (b). 
As expected, the errors in directly neighbouring
pixels are slightly anti-correlated since the effect on the shear of
adding mass density to one pixel is partially cancelled out by taking
away mass density from a neighbouring pixel. 
However the
autocorrelation function tends to zero as the pixel separations
increases, indicating that there are negligible long-wavelength
correlations in the errors.

\begin{figure}
\centerline{\epsfig{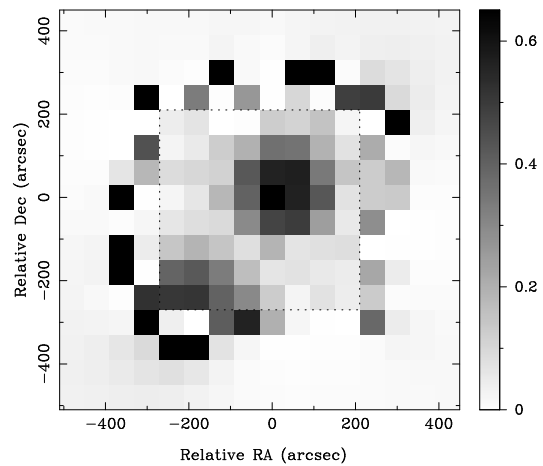}}
\caption{The maximum likelihood reconstruction of the model cluster
convergence distribution. The dotted line indicates the edge of the 
observed field.}
\label{fig9}
\end{figure}

Finally, for comparison with the MEM reconstruction shown in
Fig.~\ref{fig4}, we calculate the maximum-likelihood
(i.e. $\alpha=0$) reconstruction obtained using the same lensing data.
The resulting reconstruction is shown in Fig.~\ref{fig9}, after
160 iterations of the conjugate gradient minimisation algorithm at
which point the fractional change in $F$ per iteration is 0.1 per
cent. The reconstruction is plotted on the same greyscale as
Figs~\ref{fig2} and \ref{fig4}, 
in order to illustrate that in the observed
field the result is similar to the maximum-entropy
reconstruction. Outside the observed field, however, there are
numerous spurious features in which the convergence rises to
unrealistically large values, with a peak value of about 1.3.  Also
the condensation of mass to the bottom left of the field is not
faithfully reconstructed outside the observed box.  Moreover, on
iterating further, the peak value of the reconstructed convergence
distribution continues to rise, and after 800 iterations the
fractional change in $F$ per iteration is less than 0.001 per cent and
the peak convergence is 2.3. After an additional 800 iterations the
peak convergence rises to 4.0, although the fractional change per
iteration in the value of $F$ becomes extremely small.  For other
noise realisations, we sometimes found the maximum-likelihood solution
to be even less stable and, in some cases, the peak value of the
reconstructed convergence distribution continued to rise steadily
beyond 1500 iterations to values exceeding 7.0.  Thus, outside the
observing box we find that the maximum-likelihood reconstruction does
not, in general, converge to a numerically stable solution. 
As discussed in Section~\ref{themem}, we further note that
the maximisation of the likelihood function is performed with respect 
to $\log\kappa(j)$. Thus, the algorithm used to calculate the
maximum-likelihood reconstruction already enforces positivity 
and may therefore be compared directly with more sophisticated
reconstruction methods such as the Lucy-Richardson algorithm. We would
therefore expect such methods also to suffer from similar numerical
instabilities.

\section{Conclusions}
\label{conc}

We have presented a new maximum-entropy method for reconstructing the
projected mass distribution galaxy cluster from observations of its
gravitational lensing effects. The method is ideally suited to the
reconstruction of clusters for which the observing field is small or
irregularly-shaped and does not suffer from unwanted boundary effects
that affect the reconstructions obtained using the traditional Kaiser
\& Squires algorithm.  We find that for realistic levels of
uncertainty in the observed shear and magnification patterns, the
technique faithfully reproduces the cluster mass distribution within
the observed field and, moreover, yields a reasonable reconstruction
some distance beyond the observing box. We also show that the errors
on the reconstruction can be reliably estimated by making a Gaussian
approximation to the posterior probability distribution at its
peak. In regions where the cluster mass distribution is well
constrained by the lensing observation, we find that the estimated
reconstruction errors agree well with those obtained from Monte-Carlo
simulations.

\subsection*{ACKNOWLEDGMENTS}

We thank David Mackay for several helpful suggestions. SLB
acknowledges the PPARC for support in the form of Research Studentship.

\bsp 
\label{lastpage}
\end{document}